\documentclass[journal=jpclcd,manuscript=letter]{achemso}
\setkeys{acs}{articletitle = true}
\usepackage[version=3]{mhchem} 
\usepackage{physics}
\usepackage{xcolor}

\author{Tae Jun Yoon}
\affiliation{School of Chemical and Biological Engineering, Institute of Chemical Processes, Seoul National University, Seoul 08826, Republic of Korea}
\author{Min Young Ha}
\affiliation{School of Chemical and Biological Engineering, Institute of Chemical Processes, Seoul National University, Seoul 08826, Republic of Korea}
\author{Emanuel A. Lazar}
\affiliation{Materials Science and Engineering, University of Pennsylvania, Philadelphia, PA 19104}
\author{Won Bo Lee}
\affiliation{School of Chemical and Biological Engineering, Institute of Chemical Processes, Seoul National University, Seoul 08826, Republic of Korea}
\email{wblee@snu.ac.kr}
\author{Youn-Woo Lee}
\email{ywlee@snu.ac.kr}
\affiliation{School of Chemical and Biological Engineering, Institute of Chemical Processes, Seoul National University, Seoul 08826, Republic of Korea}

\title{Topological Characterization of Rigid-Nonrigid Transition across the Frenkel Line}

\begin{document}
\makeatletter
\setlength\acs@tocentry@height{5.08cm}
\setlength\acs@tocentry@width{5.08cm}
\makeatother
\begin{tocentry}
\begin{center}
	\includegraphics{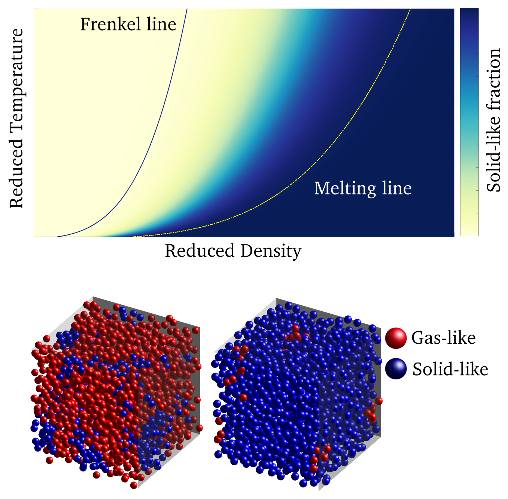}
\end{center}
\end{tocentry}
\begin{abstract}
  The dynamics of supercritical fluids, a state of matter beyond the gas-liquid critical point, changes from diffusive to oscillatory motions at high pressure. This transition is believed to occur across a locus of thermodynamic states called the Frenkel line. The Frenkel line has been extensively investigated from the viewpoint of the dynamics, but its structural meaning is still not well understood. This letter interprets the mesoscopic picture of the Frenkel line entirely based on a topological and geometrical framework. This discovery makes it possible to understand the mechanism of rigid/non-rigid transition based not on the dynamics of individual atoms, but on their instantaneous configurations. The topological classification method reveals that the percolation of solid-like structures occurs above the rigid-nonrigid crossover densities.\\\\\\
\end{abstract}

Despite its abundance in nature~\cite{bassez2003high,ingersoll1969runaway,fomin2014dynamic} and utilization in industry~\cite{reverchon2006supercritical,byrappa2007hydrothermal,byrappa2008nanoparticles}, supercritical fluid has been regarded as a \textit{terra incognita}~\cite{brunner2004supercritical,bolmatov2013evidence} of the fluid physics over a century. Its anomalous behaviors in the vicinity of the critical point~\cite{simeoni2010widom} and in the high-pressure region are not entirely understood. In the high-pressure region, the dynamics of a particle change from diffusive (gas-like) to oscillatory (solid-like) motions as the system pressure increases. To understand this dynamic crossover in the supercritical region, Brazhkin et al. first suggested the concept of the Frenkel line~\cite{brazhkin2012two}. Since then, the Frenkel line has been theoretically located based on the phonon theory~\cite{bolmatov2012phonon,bolmatov2015unified} and thermodynamic criteria~\cite{brazhkin2013liquid}. Recently, this notion of the dynamic crossover was experimentally validated using spectroscopic techniques \cite{smith2017crossover,prescher2017experimental}.
    
    Although the Frenkel line was initially defined in the dynamic context, namely, the atomistic jump time to reach its nearest neighbors, a mesoscopic interpretation of this dynamic transition is not well understood. A variety of molecular-level classification schemes \cite{heyes1988percolation,campi1999definition,yoon2018probabilistic,idrissi2013characterization,ovcharov2017particle} were proposed to understand the behavior of supercritical fluids by defining configurational clusters which are defined as a group of particles which are linked pair by pair based on a cutoff distance or a Voronoi cell volume. However, they do not yield the percolation lines linked with this dynamics transition because the Frenkel line is irrelavant to the gas-liquid criticality \cite{fomin2018dynamics,yoon2018two}. Thus, there have been a few attempts to directly relate the structural characteristics with the Frenkel line. Three such examples are the anomaly in the third maximum of pair correlation function \cite{bolmatov2013evidence}, the packing fraction of effective hard spheres \cite{fomin2014dynamic}, and tetrahedrality \cite{ryltsev2013multistage,brazhkin2014frenkel}. Bolmatov et al. reported that the height of the third maximum of the pair correlation function diminishes when the system density is below the crossover density. Fomin et al. \cite{fomin2014dynamic} and Ryltsev et al. \cite{ryltsev2013multistage} paid attention to the notion of percolation. Yet these approaches are limited in the following aspects. First, although the change of the pair correlation function was experimentally observed, there has been a controversy in its interpretation. Bolmatov et al. reported that the medium-range order persists below the crossover density based on the partial pair correlation function of $\mbox{CO}_{2}$\cite{bolmatov2014structural}. Bryk et al. deemed that no abrupt change of the pair correlation function occurs. They also criticized that the third-peak intensity of the pair correlation function is so weak that it is often difficult to distinguish from numerical errors \cite{bryk2017behavior}. Second, although the tetrahedrality and the packing fraction of effective hard spheres reach their percolation transition densities near the Frenkel line \cite{fomin2014dynamic,ryltsev2013multistage}, they explain how the solid-like structure evolves in the rigid liquid region between the Frenkel line and the melting line only approximately.
	
	In this letter, we propose a topological method to understand the mesoscopic picture of the rigid-nonrigid transition across the Frenkel line. Since the diffusive motion of a particle is governed by its local configuration, we adopted the topological framework for local structure analysis \cite{lazar2015topological} to investigate the systems. In this framework, a system composed of $N$ atoms is first partitioned into $N$ Voronoi cells, where the Voronoi cell of each atom is the region of space closer to it than to any other atom. Information about the connectivity of the edge graph of a Voronoi cell records information about the manner in which neighboring atoms are arranged relative to the central atom and to one another. This topological information can be encoded as a series of the integers called a Weinberg vector \cite{weinberg1966simple}. The list of the Weinberg vectors, hence, indicates the types of the nearest neighbors' configurations that a particle in a specific system can have. We expected that this topological framework could estimate the `rigidity' of an individual particle systematically.
	
	We first used VoroTop \cite{lazar2017vorotop} to obtain the set of Weinberg vectors of the ideal gas and the maximally random jammed (MRJ) state, which are the two opposite limits of a fluid phase regarding the particle dynamics. In contrast to the ideal gas, in which molecules freely diffuse without being affected by neighbors, the MRJ state is defined as a state where the particles are disordered while being mechanically rigid \cite{torquato2000random}. The distributions of types of the Voronoi cells from these thermodynamic limits were distinct. In the ideal gas, 9,710,780 types of the Voronoi cells were identified when fifty configurations of 500,000 particles were used. The list of the most frequently observed Voronoi cells agreed with the previous work by Lazar et al. \cite{lazar2013statistical}. In the MRJ state, 325,399 types of the Voronoi cells were obtained from fifty configurations of 500,000 particles. The list of the frequently observed Voronoi cells of the MRJ state was discrepant with that of the ideal gas. The likelihood of the most abundant Voronoi cell in the MRJ state was higher than that in the ideal gas by a factor of ten. Most of the Weinberg vectors from the MRJ state were a subset of those from the ideal gas (see the Supporting Information for the details of the Weinberg vectors obtained from these two dynamic limits). The discrepancy between the list of the Weinberg vectors led us to define the rigidity of a particle from the likelihood of its Weinberg vector to appear in the ideal gas or the MRJ state. Thus, we designed the following strategy to classify a molecule as either gas-like or solid-like (Figure 1). 
    
\begin{figure*}
	\includegraphics[width=\textwidth]{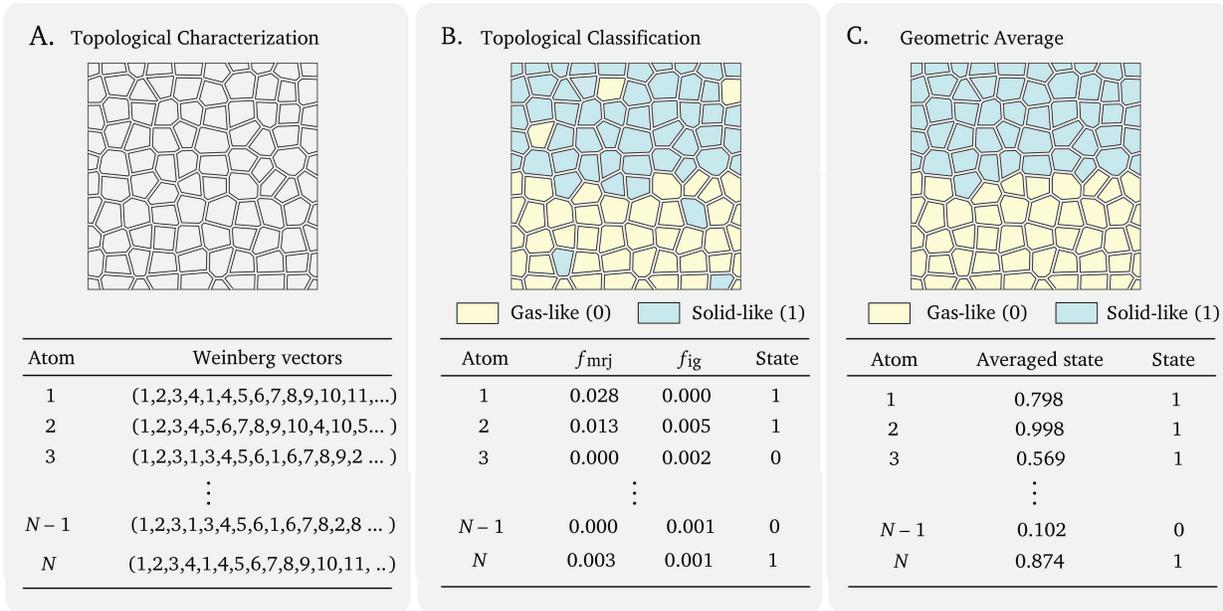}
	\caption{A two-step strategy for the topological classification of a configuration of $N$ atoms. After the topological types of the Voronoi cells are obtained, the probability of finding the type of the Voronoi cell in the ideal gas ($f_{ig}$) and that in the MRJ state ($f_{mrj}$) are compared. After the initial classification, the state number of each atom is averaged considering the first and second nearest neighbors with a weight. For visualization, two-dimensional Voronoi cells were used to represent the scheme.}
\end{figure*} 
    
    When the configuration of a system is obtained (Figure 1A), the topological information of each atom is used for the initial classification. It is labeled as a solid-like particle (state 1) when its Weinberg vector is more frequently observed in the MRJ state than in the ideal gas ($f_{mrj}>f_{ig}$). Otherwise, it is classified as gas-like (state 0).	After this topological classification, the fraction of solid-like molecules, which is defined as $\Pi_{solid}=N_{solid}/(N_{gas}+N_{solid})$, in the ideal gas was calculated as approximately 9.89 \%, and that of gas-like ones in the MRJ state was about 5.14 \%. This high fraction of solid-like molecules in the ideal gas comes from the configurational fluctuation; it causes some molecules to have the same Voronoi topology to the MRJ state by chance. Thus, the state numbers are recalculated based on the following weighted mean-field strategy. In this procedure, the state number of the $i^{th}$ molecule ($\bar{s}_{i}$) is calculated as:
	\begin{equation}
		\bar{s_{i}}=\frac{1}{N_{i}}\sum_{j=0}^{N_{i}}\left(\frac{1}{N_{j}}\sum_{k=0}^{N_{j}}s_{k}\right)
	\end{equation}
	where $s_{k}$ is the state number of the $k^{th}$ neighbor of the $i^{th}$ molecule obtained from the first step, and $N_{k}$ is the number of Voronoi neighbors of the $k^{th}$ atom. Hence, this strategy imposes a weight based on the chemical distance from the $i^{th}$ atom. When this weighted mean-field strategy is applied, the fraction of solid-like (gas-like) molecules in the ideal gas (MRJ state) decreases to zero.	
	
    \begin{figure}
	\includegraphics{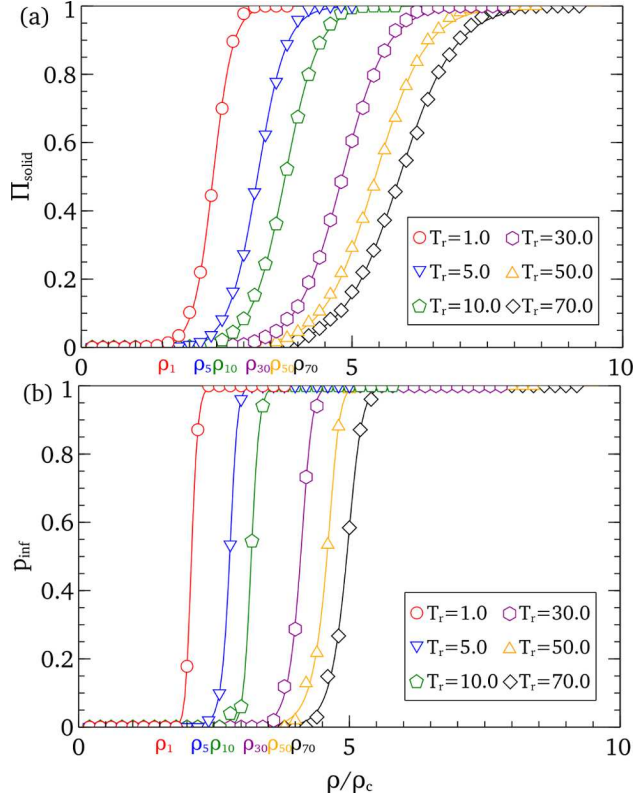}
    \caption{(a) The fraction of solid-like molecules at various reduced temperatures $T_{r}$. The solid-like fractions are well expressed by the sigmoid curves, which starts to abruptly increase near the crossover densities obtained from the two-phase thermodynamics (2PT) model \cite{yoon2018two} ($\rho_{{T_{r}}}$). The subscript $T_{r}$ indicates the reduced temperature. (b) The probabilities of finding an infinite cluster in a configuration ($p_{inf}$) at different isotherms with 2,000 molecules. They increase from zero to one in the rigid liquid region. The gradual increase of $p_{inf}$ comes from the finite size effect.}
\end{figure}

	We applied the designed topological classification method to the configurations of supercritical argon obtained from the Molecular Dynamics (MD) simulation. Figure 2 shows the fraction of solid-like molecules ($\Pi_{solid}$) at five different isotherms ($1.0\leq T_r\leq 70.0$). Regardless of the simulation temperature, $\Pi_{solid}$ showed a sigmoidal dependence on the reduced density $\rho_{r}=\rho/\rho_{c}$ that can be expressed as:
\begin{equation}
		\Pi_{solid}=\frac{1}{1+a\exp(b\rho_{r})}
\end{equation}
$\Pi_{solid}$ starts to steeply increase near the crossover densities characterized as the Frenkel line and becomes one when the system density reaches the freezing density. The solid-like fractions at the crossover densities characterized from the 2PT method \cite{yoon2018two} are almost constant ($\Pi_{solid}\sim0.02$). The influence of the finite-size effect on the classification results was negligible, which validates the robustness of the proposed method (see the Figure S1 in the Suppporting Information). 

This significant result has the following implications. First, the topological strategy designed depends on neither dynamics nor the thermodynamic properties of a system. It only classifies a molecule relying on the instantaneous configuration. Yet, the density where $\Pi_{solid}$ abruptly increases is consistent with the dynamic crossover density characterized by the phonon theory and the two-phase thermodynamics (2PT) model \cite{lin2003two,yoon2018two}. Second, the sigmoidal dependence of the solid-like fraction can be well explained by the two-state theory \cite{anisimov2018thermodynamics,yoon2018probabilistic,ha2018widom}. In the two-state theory, two interconvertible states of molecules are regarded as the following equilibrium reaction.	
	\begin{equation}
		A (gas) \rightleftharpoons A (solid)
	\end{equation}	
	where the equilibrium constant $K_{eq}$ is given as:
	\begin{equation}
		K_{eq}=\frac{\Pi_{solid}}{\Pi_{gas}}=\exp\left(-\frac{{\Delta}G^{\ddagger}}{{k_{B}T}}\right)
	\end{equation}
    Provided that the Gibbs free energy of this `reaction' linearly depends on the change of the density (${\Delta}G^\ddagger{\sim}(\rho-\rho_{c})/\rho_{c}$), Eqn. (2) is obtained from Eqn. (4). Third, as shown in our work on the supercritical gas-liquid boundary \cite{yoon2018probabilistic,ha2018widom}, the sigmoidal dependence of $\Pi_{solid}$ on the density implies that the Frenkel line is the density where the solid-like molecules percolate throughout the system. The percolation of a solid-like structure in a system could explain the appearance of the transverse excitation of positive sound dispersion (PSD) in the rigid liquid region, one of the most anomalous behavior of supercritical fluid across the Frenkel line \cite{brazhkin2012two}.

Hence, we further examined the structural characteristics of supercritical fluid across the Frenkel line. First, we attempted to show the presence of the percolation based on the finite-size scaling analysis \cite{stauffer2014introduction}. In the percolation analysis, two Voronoi neighbors, which share a face with each other, were regarded to be connected if their state numbers ($\bar{s}_{i}$) were the same. Hence, two distant molecules are assigned to a single cluster if they are connected through other atoms whose state numbers are the same as theirs. By applying the clustering algorithm by Stoll \cite{stoll1998fast}, we could analyze the percolation behavior of solid-like and gas-like structures in supercritical argon. Figure 2b shows the probability of finding an infinite cluster in a configuration ($p_{inf}$). As the density increases, $p_{inf}$ increases from zero to one above the Frenkel line. 

\begin{figure}
	\includegraphics{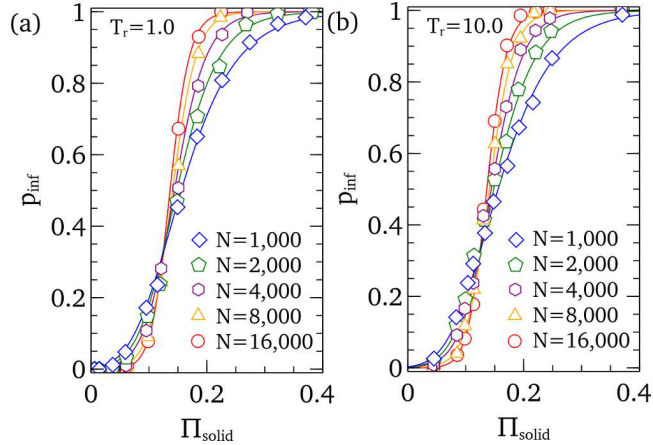}
    \caption{The influence of the system size on the percolation of solid-like clusters at (a) $T_{r}=1.0$ and (b) $T_{r}=10.0$. The spanning probabilities at different temperatures collapse to a single line when the number of molecules in a system is constant. As the number of molecules in a system increases ($N\rightarrow\infty$), $p_{inf}$ curve becomes close to a step function.}
\end{figure}

Figure 3a shows the dependence of $p_{inf}$ at $T_{r}=T/T_{c}=1.0$ on the system size and the solid-like fraction. As the number of molecules in a system increases, the slopes of $p_{inf}$ curves become steep. Likewise, $p_{inf}$ curves at $T_{r}=T/T_{c}=10.0$ in Figure 3b shows the similar dependence on the solid-like fraction. When the fraction of solid-like molecules ($\Pi_{solid}$) was used as an order parameter, $p_{inf}$ curves with the same number of molecules at different isotherms collapse to a single line. We conducted the finite-size scaling analysis \cite{stauffer2014introduction} on these systems (see the Supporting Information for the detailed procedure of the finite-size scaling analysis). The percolation threshold in the infinite system ($N\rightarrow\infty$) was obtained as $\Pi_{solid}^{c}=0.1159\pm0.0081$, which is lower than the random percolation threshold on the Voronoi lattice ($\Pi_{thr}=0.1453\pm0.002$) calculated by Jerauld \cite{jerauld1984percolation}. The correlation length exponent was $\nu=0.9030\pm0.0319$ comparable to that of the random percolation on the Voronoi lattice ($\nu=0.874\pm0.08$). Low percolation threshold implies that the percolation of solid-like structures obtained from the algorithm is a correlated percolation in which the site correlation comes from the weighted mean-field classification; the site correlation reduces the percolation threshold \cite{harter2005finite}.  

\begin{figure}
	\includegraphics{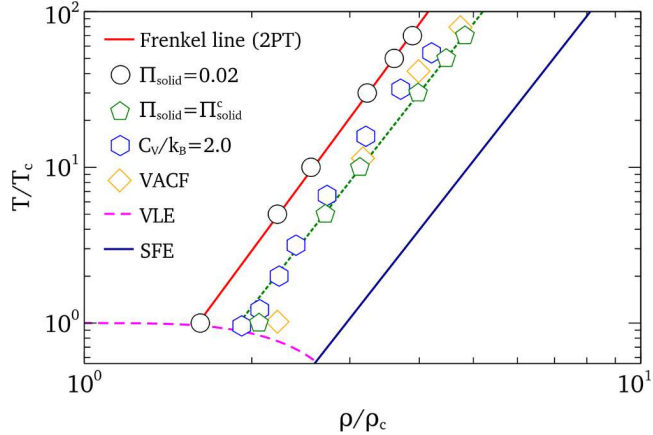}
    \caption{Crossover densities estimated from the topological classification method and other thermodynamic and dynamic criteria \cite{brazhkin2013liquid,brazhkin2012two,yoon2018two}. The first order phase transition lines (Vapor-Liquid Equilibrium line (VLE) and Solid-Fluid Equilibrium line (SFE)) are from Yoon et al. \cite{yoon2018two}}
\end{figure}

Figure 4 compares the crossover densities obtained from the 2PT model and the conventional criteria with those from the topological method. As shown in Figure 2a, the solid-like fraction abruptly increases near the Frenkel line located based on the 2PT model and the Frenkel frequency. As described earlier, the solid-like fraction is almost constant ($\Pi_{solid}=0.02$) along the dynamic crossover line. As the density increases further, $\Pi_{solid}$ becomes the percolation threshold concentration ($\Pi_{solid}^{c}$) in the rigid liquid region. The percolation densities are consistent with those from thermodynamic ($C_{v}=2.0k_{B}$) and dynamic (VACF) criteria. Hence, these results substantiate our hypothesis that the solid-like structures defined from the topological classification percolate near the crossover densities obtained from the thermodynamic and dynamic criteria. Simultaneously, it gives a physical interpretation of the discrepancy of the Frenkel lines from the Frenkel frequency and other criteria. The dynamic crossover line from the 2PT model and the Frenkel frequency is a set of onset densities where the nonrigid-rigid transition starts to occur, whereas that defined from the thermodynamic and dynamic criteria is where the solid-like structure percolates.

Our topological framework based on the Voronoi tessellation successfully interprets the physical meaning of the Frenkel line. It is a set of onset densities where the percolation of solid-like structures occurs. This result enables us to understand the physical significance of the Frenkel line as a partitioning line of the rigid-nonrigid fluids from the viewpoint of the two-state theory, which was used to explain the supercritical gas-liquid transition near the critical point and the liquid-liquid criticality. Therefore, this result deeply motivates us to conquer the \textit{terra incognita} of the fluid phase in an integrated manner.

\begin{acknowledgement}
E.A.L. gratefully acknowledges the generous support of the US NSF through Award DMR-1507013. W.B.L. and M.Y.H. acknowledge the support of the National Research Foundation of Korea Grants funded by Korean Government (NRF-2015R1A5A1036133, NRF-2017H1A2A1044355).
\end{acknowledgement}

\begin{suppinfo}
	The Supporting Information includes the detailed calculation procedures for the critical point estimation, the Molecular Dynamics (MD) simulations, the finite-size scaling analysis, the topological classification method, and the clustering analysis. Some auxiliary results in the Supporting Information also helps the readers understand the contents in the letter.
\end{suppinfo}

\bibliography{achemso-demo}
\end{document}